# Experimental tests of a model of the quantum measurement process

Alan Schaum

**Abstract**

We suggest and describe how to analyze new types of experiments that would test a proposed model of the quantum measurement process. That model produces the Born Rule as a corollary, and so agrees with conventional quantum predictions. The proposed experiments include a new type of "interrupted measurement," which would allow falsification or confirmation of the model. In order to compare it with standard quantum mechanics predictions, we also develop a new testable extrapolation of the Born Rule into the extended experimental regime.

**Background**

Reference 1 presented a general model of the "irreversible act of amplification" of a quantum state that a measurement induces. The model requires a macroscopic device as part of the definition of a measurement, meaning an apparatus with many microscopic components. The interaction of a simple quantum system (called a "target" here) with the device consists of a large number of perturbative unitary transformations of the target wave function by the apparatus' components. The evolution of the target's quantum state is modeled using any member of a certain class of stochastic processes (SPs). Each SP model acts independently of any specific detection technology but produces the Born Rule generally. Here we describe the types of calculations that would serve to falsify or confirm SP models for some notional experiments.

Each SP model describes the evolution of the magnitude squared (here called the "intensity") of any one of a (normalized) quantum state's complex amplitudes that is associated with some eigenstate of the observable O that the measurement device is designed to measure. Such an intensity $A$ is modeled as a nonnegative integer $a$ times a small positive number $\epsilon = 1/N$, with $N$ an integer. The $\epsilon$-quantization is a convenience; in the limit $\epsilon \to 0$, all testable predictions of SP models remain finite and constant.

The perturbation of an intensity
$$A = a\epsilon \qquad (1)$$

associated with any particular eigenstate of O that is caused by the $j$'th component of a measuring device is modeled as $\Delta A_j = \epsilon \Delta a_j$, with $\Delta a_j$ an integer. The evolution is serial in $j$ and for each $a$ (for any $j$) proceeds with some number of more elementary steps. These are governed by a stochastic process that generates changes of at most one in the value of $a$. With $P(a)$ the probability that intensity $a\epsilon$ reaches the value one (i.e. $a = N$), the stochastic process is defined by:

$$P(a) = p(a)\big((p(a))P(a) + (1 - p(a))P(a - 1)\big) \qquad (2)$$
$$+ (1 - p(a))\big(p(a)P(a + 1) + (1 - p(a))P(a)\big) \quad (a \neq 0, N).$$

The transition probabilities $p(a)$ plays two roles. It is the probability that intensity $a\epsilon$ donates $\epsilon$ to a temporary pool of intensities; it is also the probability that intensity $a\epsilon$ then recovers $\epsilon$ (and hence doesn't change in value). Similarly, $(1 - p(a))$ is the probability that some other intensity associated with eigenstates of O donates $\epsilon$ to the pool. Thus $p(a)(1 - p(a))$ is the probability both that $a\epsilon$ loses $\epsilon$ or gains $\epsilon$ in the process. $(1 - p(a))(1 - p(a))$ is the probability that some other eigenstate contributes to the pool as well as that some other eigenstate then draws $\epsilon$ from the pool. $p(a)$ is required to satisfy

$$p(0) = 0, \text{ and } p(a) \neq 0 \text{ if } a \neq 0. \qquad (3)$$

(A canonical example: $p(a) = a\epsilon$.) The equality in Expression 3 is required by the general model, because should an intensity reach zero, it has no $\epsilon$ to contribute to the pool. The inequality insures that a nonzero intensity cannot defect from the evolution process, an expected property of general quantum interactions. As shown in ref. [1], such SP models imply a stopping rule: Should $a$ reach either value 0 or $N = 1/\epsilon$, it stays there, the latter implying that the target state has been transformed by the measurement device into the eigenstate associated with $a$.

Notice that the probability in Equation 2 that $a$ does not change is $\big(p_M(a)\big)^2 + \big(1 - p_M(a)\big)^2$. Also, if it does change then the probabilities of $a$ increasing or decreasing by one are identical: $p(a)(1 - p(a))$. Thus, whenever $a_i$ changes to $a_{i+1}$, it executes a simple random walk in one-dimension. As proven in [1], these SP models all predict that $a_i$ reaches the value $N$, i.e., intensity reaches the value 1, with probability $A_i = \epsilon a_i$. This holds at any stage of the evolution process including, in particular, at its initial intensity $A_0 = \epsilon a_0$; this is the Born Rule.

Here we re-derive the Born Rule using "Doob's Optional Stopping Theorem" [2], which will prove useful again later for proving a new result. The theorem applies to bounded Martingale processes (such as the random walk described by the sequence $\{a_i\}$) that obey a stopping rule (such as implied by SP models). The theorem states that for such processes the expected value $E$ of $a_T$ is $a_0$, where $T$ (itself a random variable) is the stopping time. Because in SP models $\{a_i\}$ terminates at either value 0 or $N$, the theorem applies and insures that the expected value satisfies $a_0 = E(a_T) = 0P(a_T = 0) + NP(a_T = N)$. That is

$$P(a_T = N) = a_0/N = A_0. \qquad (4)$$

This is the Born rule, because $P(a_T = N)$ is the probability that the intensity $A$ has reached the value one, i.e., that the measurement has produced the eigenstate associated with intensity $A$.

**Notional experiments**

The presence of falsifiable predictions distinguishes physics from philosophy. And because SP models agree with the conventional quantum mechanics (CQM) model for all typical measurements, any test that might distinguish the two models must involve a new type of measurement, involving an interruption of the putative stochastic process posited by SP models. Such a test will also require an extrapolation of CQM for interpreting the new measurements.

We use a notation motivated by Feynman. The symbol $\{-\}$ denotes his idealized "modified Stern-Gerlach (MSG) apparatus" [3]. The MSG consists of three internal Stern-Gerlach devices designed here to act on target particles with spin-½ (such as the silver atoms of the original Stern-Gerlach experiment [4]), accepting them with nearly zero speed and transmitting them also with nearly zero speed, while recombining the internally diverging beams into a single stream. A related MSG device represented by the symbol $\{\frac{|}{-}\}$ contains a barrier that absorbs and counts all targets whose spin point "up" (in the +z direction) but transmits all targets with spin-down.

In CQM a target in a superposition of up and down states and whose spin-up intensity is $A$ is absorbed by a device of the second type with probability $A$, and the spin-down component is transmitted with probability $1-A$. A device of the first type passes any target state unchanged. We are interested here in expanding the meaning of the second symbol to represent a device (See Figure 1) whose absorbing barrier can have a variable thickness and hence, absorbance $f$ (of spin-up components). We call its action an "interrupted measurement" when $0 < f < 1$. The barrier can be calibrated by a measurement of the fraction $1-f$ of spin-up targets detected in the second MSG device (D2) in Figure 1.[1] We define the fraction of all targets absorbed by D2 as "the measurement" in the following experiments.

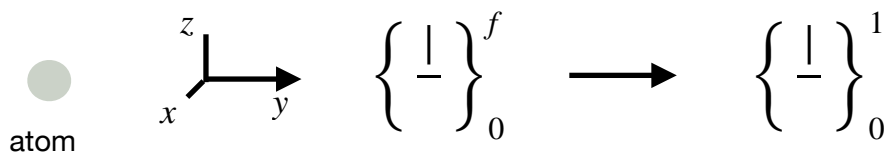

Figure 1. Spin-½ silver atom targets enter from the left, first encounter an MSG device (D1) containing an absorber of a fraction $f$ of spin-up particles. Any targets not absorbed are measured by a second, traditional MSG device (D2) that absorbs all spin-up targets and counts them. The subscript "0" indicates that a device's "spin-up direction" makes an angle $\phi = 0$ radians with the vertical.

---

[1] $f$ is implicitly defined by the $1-f$ measurement, so that other barrier actions beside absorption (e.g. scattering) that might prevent transmission are included in the meaning of $f$. Note also that for thin barriers, including a counter in the first MSG device (D1) might prove impractical.

For targets with spin in the z-x plane (y is the direction through the MSG devices) making an angle $\theta$ with the z-axis, the value $A$ of the target's spin-up ($+z$) quantum intensity is given by [3]

$$A = cos^2(\theta/2) . \tag{5}$$

In the SP model the initial intensity value $A_0$ of a target traversing and interacting with an MSG device changes repeatedly and randomly, ending in the value one (spin in $+z$ direction) with probability $A_0$ and zero (spin -z) with probability $1 - A_0$ if enough interactions occur, that is, if the absorber is thick enough to insure 100% efficiency. This requires $f > 0$ at a minimum.

The stochastic variable $A = a\epsilon$ executes a random walk in one dimension as the target interacts with D1, with small step size $\epsilon$. If $0 < f < 1$ and the number $n$ of the target's interactions with the absorber is large enough, but not too large, the probability distribution of $A$ can be approximated with a Gaussian. Figure 2 shows the result for $A_0 = .5$ and standard deviation $\epsilon\sqrt{n} = 0.15$. For greater values of $n$ or different values of $A_0$, the deviation of the distribution from Gaussian becomes pronounced.

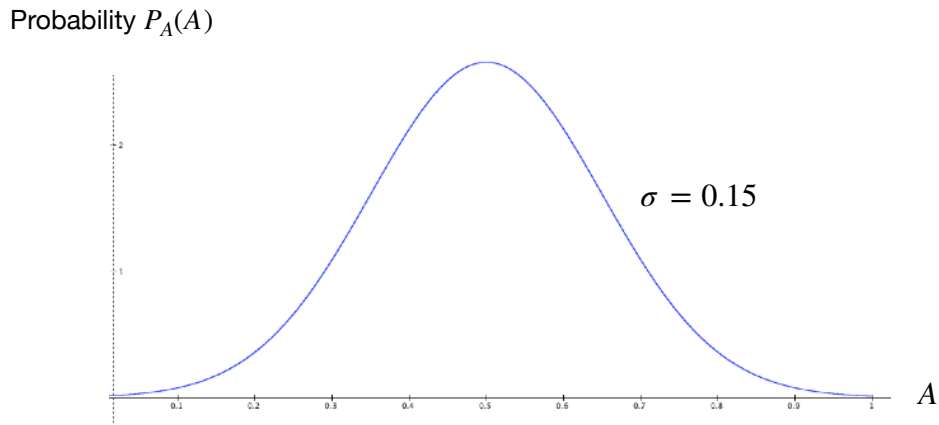

Figure 2. Distribution of +z intensities $A$, with $A_0 = 1/2$, after an interrupted measurement too abbreviated for $\{A_i\}$ to terminate at zero or one.

Figure 3 shows the qualitative behavior for $A_0 = 2/3$, including a $\delta$-function at one endpoint for those instantiations of the random variable $A$ that have walked to the value one (and hence remain there), corresponding to the $+1/2$ eigenstate of z-component spin. The true distribution has no known closed form. Its representation is difficult to compute because, for example, the stopping points at $a = 0, N$ prevent the random walk that $A$ executes from crossing (and recrossing) either boundary. If, by contrast, $n$ is large enough to produce a traditional measurement, then the probability distribution devolves into a pair of $\delta$-functions: $P_A(A) = (1/3)\delta(A) + (2/3)\delta(A - 1)$.

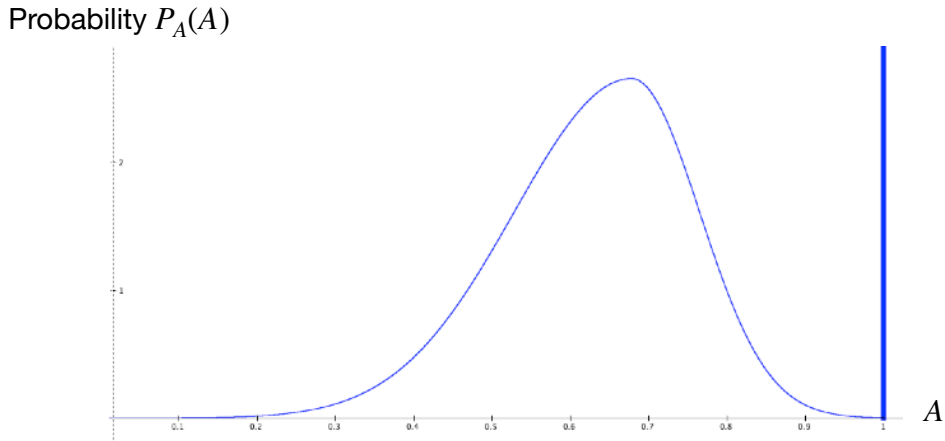

Figure 3. Qualitative distribution form of +z intensities with $A_0 = 2/3$ in which some terminations occur at $A = 1$.

## Measurement predictions of the Stochastic Process (SP) model

Next we explore some experimental tests of the SP model for spin-½ quantum targets.

### z-direction measurements

Each realized value of $A$ depicted in Figure 2 represents the spin-up ($+z$) intensity some target attained after some $n$ interactions with components of the first ("interrupted measurement") device D1 of Figure 1. No value reached the limiting pure spin-up $A = 1$ possibility, so that D1 makes no detections, and $f = 0$. All targets proceed to the measurement device D2, which then converts any target with intensity $A$ into the spin-up state with probability $A$, so that the mean fraction $\mu$ of total spin-up conversions is

$$\mu = \int_0^1 A P_A(A) dA = \epsilon \int_0^N a P(a) da, \tag{6}$$

which can be evaluated, again using Doob's Optional Stopping Theorem.

The Martingale process described by $a_0, \ldots, a_n$ satisfies the criterion of the theorem, because it includes a definite maximum stopping time after at most $n$ steps. Thus, the mean of $a$ is $a_0$, and the mean of $A$ is $A_0$. Therefore, a mean fraction $A_0$ (the initial target intensity and the mean of $P_A$) of targets are converted into a spin-up state by D2. The is the same result according to CQM with $f = 0$, and so achieving a condition described by Figure 2 might not suffice to distinguish CQM from SP models.

If, on the other hand, a distribution like that of Figure 3 applies, the mean value of $A$ is again $A_0$ (Doob's theorem still applies). But now the scale $s$ of the $\delta$-function at $A = 1$ represents the fraction of target intensities converted to the value one, a mean fraction $f$ of which is absorbed by D1. Thus D2 measures the fraction $A_0 - fs$. This can be compared with the CQM prediction for the same D2 measurement: $(1 - f)A_0$, as will be shown.

We expect the entire range of both $f$ ([0, 1]) and $s$ ([0, $A_0$]) to be attainable as $n$ increases from zero without limit. However, we frame no hypotheses about the relationship of $s$ to $f$. For example, no assumption of the SP model disallows the possibility that for any measurable nonzero $f$, $n$ must be so large that s has already reached its maximum value of $A_0$. In this case, SP and CQM models agree (D2 measures $(1 - f)A_0$), and therefore z-direction measurements might not suffice to distinguish the model for any value of $n$.

It will be shown that the CQM output in these hypothetical experiments consists of at most three distinct pure quantum states, while the SP model can produce many intermediate quantum states. This implies that the density matrix descriptions of the outputs of CQM and SP differ. Therefore, measurable differences in the predictions of the two models must exist.

### x-direction measurements

More discriminating tests can be constructed by rotating the measurement device. Assume now that D2 is rotated by an angle $\phi = 90°$ (See Figure 4) around the $y$-axis, so that it completely absorbs targets with spin in the $+x$ direction. If an experimenter can adjust the absorber to restrict the number of interactions $n$ (and hence the size of $\sigma$) to correspond to a distribution as depicted in Figure 2, or one even narrower (with smaller $n$), then the Gaussian assumption can suffice in estimating the effects of an interrupted measurement. (Again, this corresponds to zero conversions to $A = 1$, so no absorption occurs in D1.) With that restriction, we now assume that the input spin-½ particles are polarized in the $+x$ direction ($\theta_0 = \pi/2$ in Equation 5) before encountering D1. Let $B = b\epsilon$ represent the corresponding $+x$ intensity, so that the initial spin component in the $+x$ direction, $B_0 = 1$. Then from Equation 5, $A_0 = 1/2$.

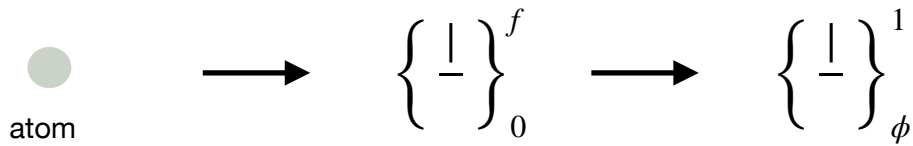

Figure 4. In general, the "measurement" can be made by a modified Stern-Gerlach device rotated an angle $\phi$ from vertical.

Analogously to Equation 5, the evolution of $B$ as a target interacts with D1 corresponds to an evolution of $A$, according to

$$B = cos^2((\pi/2 - \theta)/2) = \left(\frac{cos\theta/2}{\sqrt{2}} + \frac{sin\theta/2}{\sqrt{2}}\right)^2 = 1/2 + sin(\theta/2)cos(\theta/2), \quad (7)$$

that is (from Equation 5)

$$B = 1/2 + \sqrt{A(1-A)}. \quad (8)$$

Similarly,

$$A = 1/2 + \sqrt{B(1-B)}. \quad (9)$$

The + sign is appropriate when $B_0 > 1/2$, because if the evolution of $A$ from the initial value $A_0 = 1/2$ ever reaches either zero or one, the z-spin evolution process in D1 stops, according to the SP model. Therefore, any evolution of the pair $(A, B)$ caused by the first device is confined to the upper semicircle shown in Figure 5, and so $B$ can never reach a value less than 1/2.

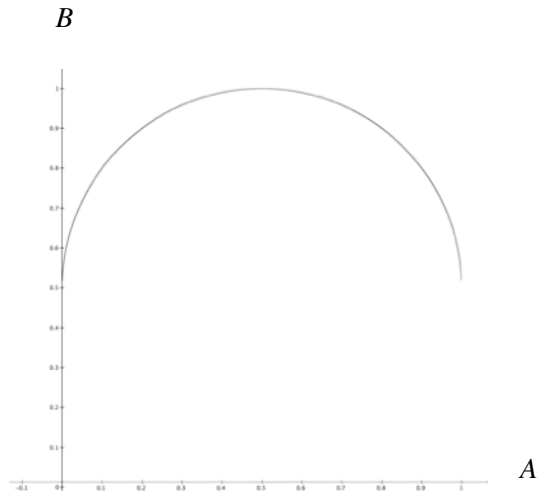

Figure 5. Relation between $+z$ and $+x$ spin coordinates.

As $A$ executes a random walk (of step-size $\epsilon$) in the interaction of the target with D1, the mean value of $B$ is the mean fraction of "up" measurements recorded by the rotated, measurement device (i.e., in the $+x$ direction). Finding this value experimentally would constitute a preliminary test of the SP model. Notice that Doob's Theorem cannot be applied to $B_i$; the process $\{b_i\}$ is not a Martingale.

If the variation of $A$ corresponds to Figure 2 (or with smaller variance), then the $A$ probability distribution approximately satisfies

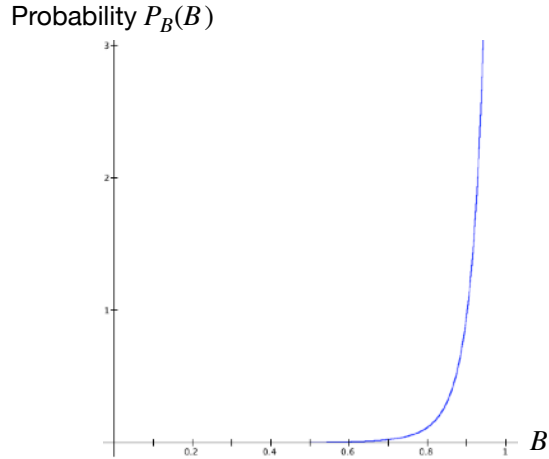

Figure 6. Distribution of $+x$ spin distribution generated by the $A$-distribution of Figure 2.

$$p_A(A) = (1/\sigma\sqrt{2\pi})\exp\left(-\frac{1}{2\sigma^2}(A-1/2)^2\right), \tag{10}$$

in which the value of $\sigma$ (and hence $n$) is determined by the experimenter's chosen barrier thickness.

The distribution of $B$ is given by

$$p_B(B) = \left|\frac{dA}{dB}\right|p_A(A) = \frac{2B-1}{2(B(1-B))^{1/2}}(1/\sigma\sqrt{2\pi})\exp\left(-\frac{1}{2\sigma^2}(B)(1-B)\right), \tag{11}$$

in which Equation 10 has been used and Equation 9 has been differentiated. This $B$-distribution, corresponding to the $A$-distribution of Figure 2, is plotted in Figure 6.

In the SP model the mean value $\mu$ of $B$ predicts the fraction of $+x$ detections made by the rotated measurement device. It is given by

$$\begin{aligned}\mu &= \int_{1/2}^{1} B p_B(B) dB = \int_0^1 \left(1/2 + \sqrt{A(1-A)}\right) p_A(A) dA \\ &\approx \int_{-\infty}^{\infty} \frac{(1/2 + \sqrt{A(1-A)})}{\sigma\sqrt{2\pi}} \exp\left(-\frac{1}{2\sigma^2}(A-1/2)^2\right) dA,\end{aligned} \tag{12}$$

in which Equations 8 and 10 have been used, as well as the small size of $\sigma$. The computed value of $\mu$ for a mean $A$ of ½ and variances no greater than that in Figure 2 is no more than a few percent less than one. A value near one also can be expected in CQM. The absence of $\delta$-functions in Figure 2 means that, with little absorption, CQM dictates that the corresponding $\mu_{CQM} \approx 1$, as well. Better tests to distinguish SP from CQM models would require larger values of $\sigma$, thicker absorbers.

Notice that for very large $\sigma$ (corresponding to many interactions and a traditional measurement) the above approximations do not hold. The distribution of Figure 2 is replaced by a pair of symmetric $\delta$-functions at the endpoints, and the presence of one at $A = 1$ means that half the atoms are converted to intensity one. If $n$ is large enough, then $f = 1$, so that half are stopped by D1. Of the remaining half corresponding to $A = 0$, half of these are detected by the (rotated) measurement device, for a total fraction ¼ measured. On the other hand, for very small $\sigma$, it follows that $f \to 0$ and $P(B) \to \delta(B - 1)$. D1 is transparent, and all targets are detected by the measurement device. In both these limits the SP results agree with those of CQM.

In an experimental test of these ideas, intermediate values of $\mu$ will result from larger values of $\sigma$, –and should be most useful–but calculating them requires a more accurate characterization of $p_B(B)$ than that depicted in Figure 3 starting with, for example, the discrete form

$$P_A(A) = P_A(m\epsilon) = 2^{-n} \binom{n}{\frac{m+n}{2}} \tag{13}$$

(( ) is the binomial coefficient), which is valid when the number $n$ of steps is small enough that the value of $a$ has never reached either 0 or $N$ in executing its random walk. For larger values of $n$, a Monte Carlo simulation of the exact discrete stochastic process would be called for, in which the possible accumulation of probabilities at the end points, $A = 0, 1$ is included in the model-ing. Of course, $n$ (and hence $\sigma$) is only indirectly under the control of the experimenter, through control of the barrier's thickness and the value chosen for $\epsilon$.

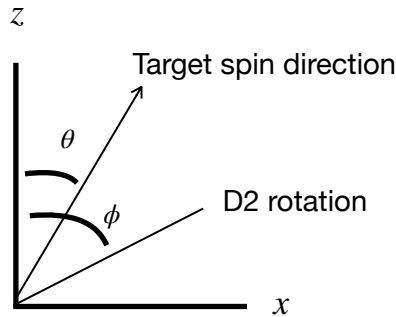

Figure 7. Geometry of the measurement system.

## General rotations

For completeness, we now consider preliminary results for arbitrary rotations $\phi$ (see Figure 4) of the measurement device (and for $B_0 \geq 1/2$). (The above example corresponds to $B_0 = 1$ and a measurement device rotated by $\phi = 90°$.) Let $B$ represent the intensity in the coordinate system of the rotated (now by angle $\phi$) device. The more generalized version of Equation 7 (see Figure 7) is

$$B = cos^2\left((\phi - \theta)/2\right) = Acos^2(\phi/2) + (1-A)sin^2(\phi/2) + sin\phi\sqrt{A(1-A)}\,. \qquad (14)$$

As an example, Figure 8 relates $B$ to $A$ (Equation 14) for the particular value $\phi = \pi/4$. Calculating the mean number of spin-up (in the $\phi$-rotated coordinate system) measurements can proceed along lines analogous to the $\phi = \pi/2$ case above. The range of $\sigma$ values where a simulation is not necessary is now smaller than the earlier value of 0.15. Monte Carlo methods are essential for exploring the full range of parameter values.

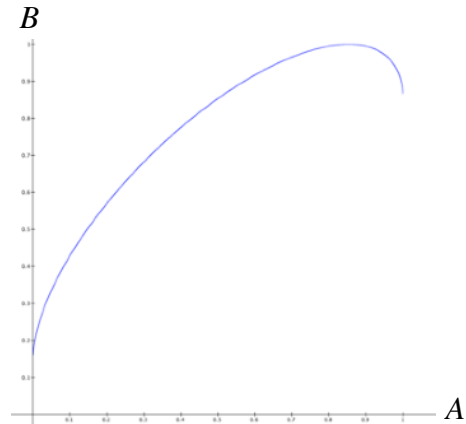

Figure 8. $\phi = \pi/4$ coordinate ($B$)
vs $\phi = 0$ coordinate ($A$)

## Model for conventional quantum mechanics (CQM)

Falsification or confirmation of the SP model would be demonstrated, if its predictions conflicted with those of CQM. It does not conflict in conventional experiments. However, it is not clear what the CQM predictions are for the kind of "interrupted" measurements described above.

Therefore, we suggest an extrapolation. We are interested in representing in standard quantum mechanics terms the results of sending a spin-½ target in any pure quantum state through any single MSG device, now to include $0 < f < 1$.

We know that for $f = 1$, the D1 device in Figure 1 converts a general spin-½ quantum state into either of two quantum states, spin-up, which is absorbed, or spin-down, which is transmitted. We also know that for $f = 0$ the device transmits the initial state unchanged. It seems simplest then to assume for CQM that a device operating with $0 < f < 1$ converts a pure quantum state into one of three possible types of pure state. To wit, if states spin-up, spin-down, or the original state are produced (after many experiments) with probabilities

$$fA_0 \ (absorbed), \ f(1 - A_0), \ 1 - f , \qquad (15)$$

respectively, then both the limiting cases ($f = 0, 1$) are consistent with known results.

Furthermore, if the initial state is spin-up, then $A_0 = 1$, and this model predicts at the output of D1 no $-z$ component, but in addition to the absorbed fraction $f$, a transmitted fraction $1 - f$ of spin $+z$, as expected. Similarly, if the initial state is spin-down, then $A_0 = 0$, and no spin-up is transmitted, while spin-down is transmitted via two virtual routes, one with probability $f$, the other $1 - f$; that is, the spin-down initial state is transmitted faithfully through D1. Thus, all obvious sanity checks are confirmed. The probabilities in (15) constitutes our CQM model for D1-type "interrupted" measurements.

This CQM model of what such an MSG device produces appears to be new. We are unaware of an any other attempt to describe the effect of this type of interrupted measurement process, by which is meant a weak enough interaction of a target system with a measuring device that it might not quite qualify as a measurement.

### CQM measurement predictions

Recall that the SP model predicts, for $\phi = 0$, that the mean fraction of spin-up targets produced by both devices in the Figure 1 configuration is $A_0$. The model in Expression (15) predicts a mean fraction of spin-up conversions by the combined MSG devices of

$$fA_0 + (1 - A_0)(0) + (1 - f)A_0 = A_0 , \qquad (16)$$

identically to the SP prediction, as claimed earlier.

Choosing $\phi \neq 0$ allows greater opportunity to explore differences in the models. The spin-up intensity (relative to the measurement device) is given by Equation 14. According to the CQM model expressed by the probabilities in (15), the fraction $fA_0$ of targets is removed by D1. For the parts transmitted:

(1) The probability that the spin-down state is produced by the interrupted measurement is $f(1 - A_0)$ and, as in Equation 5, its measured component in $+z$ direction of D2 is $f(1 - A_0)cos^2((\pi - \phi)/2) = f(1 - A_0)sin^2(\phi/2)$ (see Figure 7);
(2) The probability that the original state survives is $1 - f$, and its component in the spin-up direction of D2 is $(1 - f)B$, with $B$ given by Equation 14.

The sum of the two probabilities, which constitute the mean fraction $F$ of spin-up measurements reported by D2, simplifies to (see Figure 9):

$$F = (1 - f)A_0 cos^2(\phi/2) + (1 - A_0)sin^2(\phi/2) + (1 - f)sin(\phi)\sqrt{A_0(1 - A_0)}. \qquad (17)$$

Thus, experimental results can be tested easily against a closed-form prediction. Confirmation of Equation 17 for a wide range of parameters, $f$, $\theta$, $\phi$, would disprove the SP model.

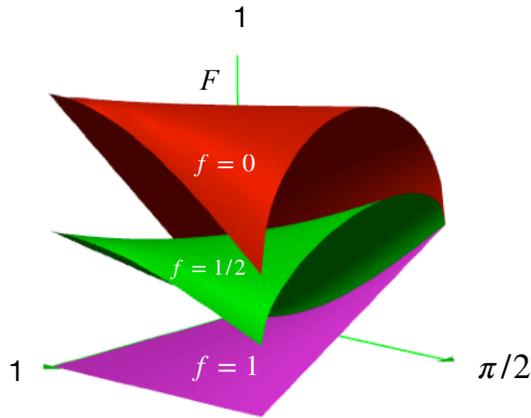

Figure 9. The fraction $F$ of spin-up detections recorded by D2 of Figure 1, as a function of the initial spin-up intensity $A_0$ and the device's angle of rotation $\phi$ relative to the vertical (Equation 17).

The key to distinguishing SP and CQM model predictions lies in choosing a value of $n$ in the Monte Carlo simulation that is small enough that a substantial continuous part of $P_A(A)$ persists. The more conversions are made to $\delta$-functions, the closer the results will approximate standard quantum results.

**Discussion and Summary**

We have described two testable models of the quantum measurement process: the Stochastic Process model and an extrapolation of the Conventional Quantum Mechanics interpretation. We showed how both could be investigated using idealized Stern-Gerlach-type devices incorporating a new type of "interrupted" measurement.

The extrapolation of CQM, described in (15), is a hypothesized interpretation of how to apply conventional quantum methods to the experimental conditions implied by the new type of measurement. It is also easily falsifiable through experimentation, with closed-form predictions available (Equation 17).

We also described the (more difficult) type of calculations necessary for exploring some experimental consequences of the SP model. For a very limited range of parameter values (all of which require $n > 0$ and $f \approx 0$) numerical integration suffices. A more comprehensive study would require more extensive, Monte-Carlo-type computer simulations for a variety of parameter values: $\epsilon, n, \phi, A_0$, but especially $f$. It is conceivable that for even small values of $f$, $n$ will be found large enough that the D1 measurement is complete, not interrupted. This would result in the standard CQM result. Consequently, the feasibility of experiments suggested here using imperfect absorbers might depend on access to very thin, perhaps mono-atomic material layers, e.g., graphene or goldene [5]. However, other types of measurement could be devised to compare the SP and CQM models.

It has been pointed out [6] that a measurement [7] has been made that demonstrated the (effective) double-slit diffraction of atoms. Laser light was also directed at the slits, destroying the interference pattern. With similar equipment, experiments analogous to those with MSG devices described above could be undertaken, in principle. The screen detecting the interference pattern plays role of D2, the double slit plays role of D1, and the laser intensity plays the role of imperfect absorber (especially if it can be focused on only one of the slits), making the "interrupted" measurement. Decreasing the intensity of the laser light (not done with the original experiments) would be analogous to decreasing the absorbance $f$ in the above experiments. Repeating such experiments while varying laser intensity might test these measurement models more easily.

Other measurements analogous to those described here are described by Rovelli [8], with photons replacing spin-½ particles and gratings replacing Stern-Gerlach devices. The absorber (Rovelli's hand) in those experiments could be replaced with an optical filter with nonzero transmittance, producing the same type of interrupted measurement discussed here. Undoubtedly many other types particular phenomenologies, especially among those used to investigate quantum foundations [9], could be exploited to investigate these models.

It is hoped that by illustrating principles of the SP and CQM models with idealized devices, this paper might serve as a useful guide for analyzing and designing any realizable experiments one might devise to test these hypotheses about the nature of the quantum measurement process and the source of the Born Rule. Experimental tests of the SP model could finally at least point to an answer to the question: Where does the quantum world end and the classical begin?

Schrödinger once opined that quantum mechanics "leaves unanswered the question of how a measurement on the system causes the vast evolving catalogue (of possible eigenvalues) to collapse to a single indelible entry. That is enough to convict quantum mechanics of incompleteness." [10] If the SP model can be confirmed experimentally–or even if only the CQM model can be falsified–perhaps one step toward completeness will have been taken.